\documentclass[
    reprint,
    amsmath,amssymb,
]{revtex4-2}

\usepackage{graphicx}
\usepackage{dcolumn}
\usepackage[dvipsnames]{xcolor}
\usepackage{bm}
\usepackage{hyperref}
\usepackage{url}
\usepackage{upgreek}

\hypersetup{
    colorlinks=true,
    linkcolor=blue,
    filecolor=magenta,      
    urlcolor=cyan,
    pdftitle={Overleaf Example},
    pdfpagemode=FullScreen,
    }
    
\begin{document}
\preprint{APS/123-QED}

\title{Temporal bandwidth of consecutive polariton condensation}

\author{Mikhail Misko$^{1,\bullet}$}
\author{Anton D. Putintsev$^{1,\bullet}$}
\thanks{Contact author: anton.putintsev@skoltech.ru}
\author{Denis Sannikov$^{1}$}
\author{Anton V. Zasedatelev$^{1}$}
\author{Ullrich Scherf$^{2}$}
\author{Pavlos G. Lagoudakis$^{1}$}
\thanks{Contact author: p.lagoudakis@skol.tech\\$^{\bullet}$ M.M. and A.D.P. contributed equally to this work.}
%\footmark{abc}

\affiliation{$^1$Hybrid Photonics Laboratory, Skolkovo Institute of Science and Technology, Territory of Innovation Center Skolkovo, Bolshoy Boulevard 30, building 1, 121205 Moscow, Russia}
\affiliation{$^2$Macromolecular Chemistry Group and Institute for Polymer Technology, Bergische Universität Wuppertal, 42119 Wuppertal, Germany}

\date{\today}

\begin{abstract}

The advent of organic polaritonics has led to the realisation of all-optical transistors, logic gates, and single photon-switches operating at room temperature. In this Letter, we develop a microscopic theory accounting for thermalisation, vibron-relaxation, and kinetic bimolecular quenching of polaritons and excitons to investigate the intrinsic limitations of the temporal separation of consecutive polariton condensates. We test and verify our theoretical predictions using an optical pump-pump configuration with different pulse widths and unravel the importance of kinetic losses in defining the upper limit of the temporal bandwidth, reaching  $\sim 240~\rm GHz$.

\end{abstract}
                              
\maketitle

The breakdown of Dennard scaling law \cite{Fiori2014} has limited the computational capabilities of modern processors at few GHz. Different optical platforms have been proposed to overcome this limit \cite{Jose2022, Shuo2018, Miller2017, Volz2012} with the promise of accelerating classical computing using optical neural networks \cite{SturmONN2022}, reducing losses of photonic integrated circuits \cite{Xu2021,Shen2017}, and achieving ultra-low energy-per-bit costs of information processing \cite{deCea2021}. However, the scalability of their operational functionalities is bounded by the large volume of the nonlinear media \cite{Miller2010, Caulfield2007, Hardy07}. Furthermore, the operation of a NOT gate, necessary for full logic functionality, requires coherent control in order to switch-off an optical signal with another \cite{Kundermann2003,Miller2010com}. An alternative platform based on exciton-polariton condensates offers large intrinsic nonlinearities through their matter component \cite{Kavokin2017}, broad spectral tunability \cite{Putintsev2023,McGhee2023,Sannikov2019,Tsotsis2014}, and the potential for electrical injection \cite{Schneider2013, Jayaprakash2019, Christogiannis2013, Paschos2017}. Recent advances of the exciton-polariton physics in organic semiconductor microcavities have translated all-optical transistor functionalities from cryogenic temperatures \cite{hwang2009,ballarini2013,chen2013} to room temperature operation \cite{Zasedatelev2019}, allowing for the realisation of the universal NOR gate without the necessity for coherent control and critical biasing \cite{Sannikov2024} and single-photon switching \cite{Zasedatelev2021}.  

\begin{figure}[t!]
    \includegraphics[scale=0.047]{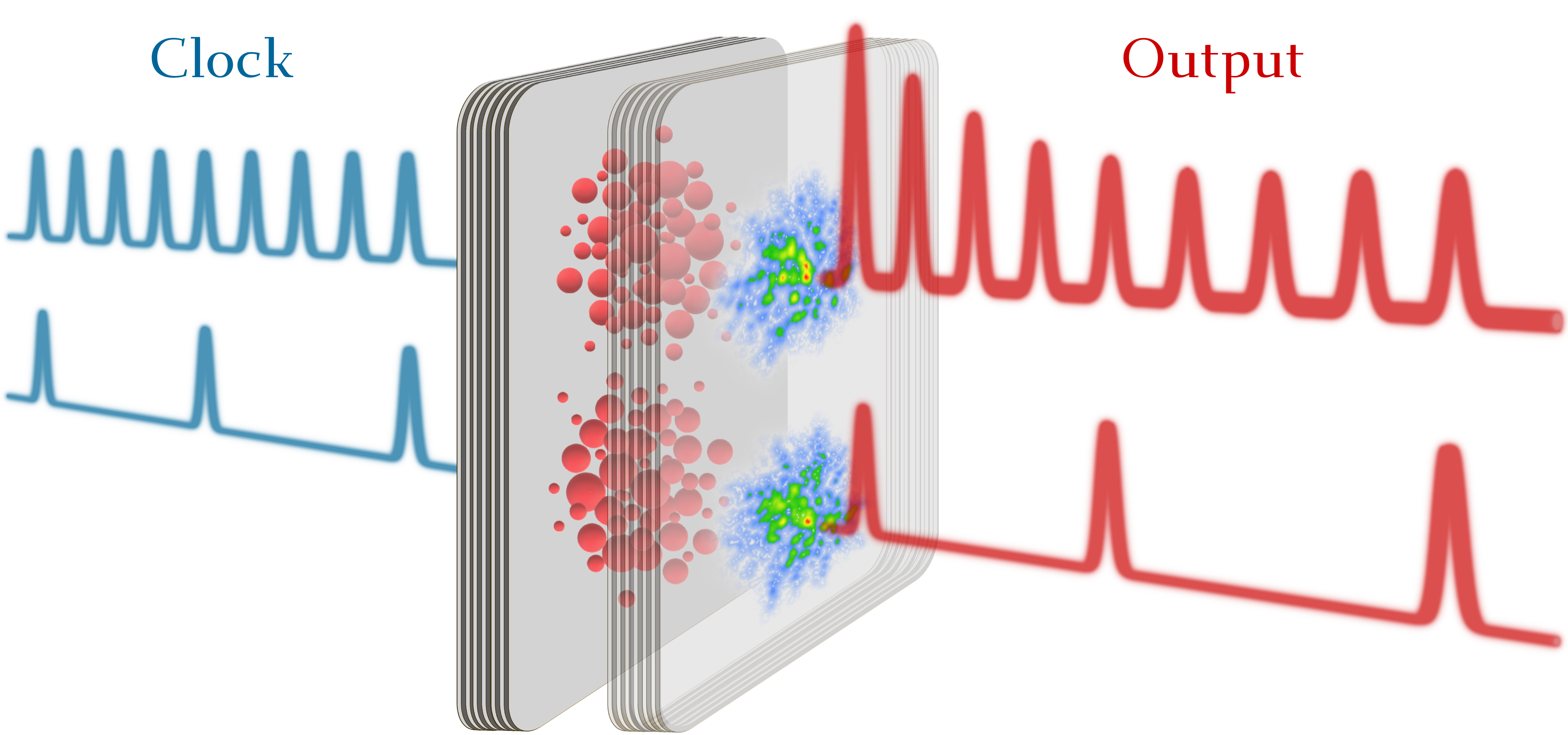}
    \vspace*{-0.25cm} 
    \caption{\label{fig:1} Schematic of semiconductor microcavity as an I\slash O logic gate. Non-resonant pump-pulse trains at slower\slash faster frequency acting as a transistor clock (blue colored pulses). The blue pulses inject a reservoir of hot excitons that energetically relax to form a polariton condensate (red bubbles). The polariton emission occurs in light bursts (red colour pulses; real-space single-shot emission profiles are overlaid on the top mirror). The lower case illustrates normal operation of the transistor output: each bit of the signal is executed independently for equal pump-bias resulting in equal outputs. The upper case is above the critical-frequency regime, wherein each consequent pump-bias adds to the preoccupied polariton condensation resulting in a ``pile-up" effect of consecutive bits in the output signal.}      
\vspace*{-0.5cm} 
\end{figure}

The limitation on the operation bandwidth of a polariton transistor relates to the temporal dynamics of a polariton condensate. Essentially, for a polariton-based binary logic two distinctively different polariton occupation densities are required to separate between logic-states ``1" (high) and ``0" (low). The transition between those defines the concept of executing consecutive operations in logic gates. Their operational speed in turn is defined by the slowest of operations that is the execution of consecutive logic-states of ``1”. Their independent realisation requires that the system should be sufficiently drained of the residual polariton population from each ``1" state. The level of residual polariton population is of particular importance due to the strong amplification observed even under incoherent seeding of the polariton ground state \cite{Zasedatelev2021}; the latter prevents the proper distinction between two consecutive logic-states. Therefore, we expect that by gradually increasing the operational frequency (decreasing the temporal separation between the two pulses) we will reach the amplification regime when the residual polariton population from the first pulse serves as the ``seed" for the amplification for the second pulse. Since polariton condensation requires some time to occur, the maximum of such amplification is expected to be at some non-zero time-delay between the pulses. Figure \ref{fig:1} depicts the proposed concept: an organic microcavity acting as an Input\slash Output logic gate allows for the formation of a polariton condensate under non-resonant pulsed optical excitation, or clock-pumping (blue colour). The output photoluminescence (PL) (red colour) strongly depends on the time-delay between the input pulses. The bottom row in Fig.\ref{fig:1} represents a normal operation without the ``pile-up" effect in the output signal, whilst the top row demonstrates the high clock-rate with an abrupt amplification of the output signal when the residual polariton population of the ground state from the previous pulse is not sufficiently depleted before the arrival of the next pulse.

\begin{figure}[t!]
    \centering
    \includegraphics[scale=1.15]{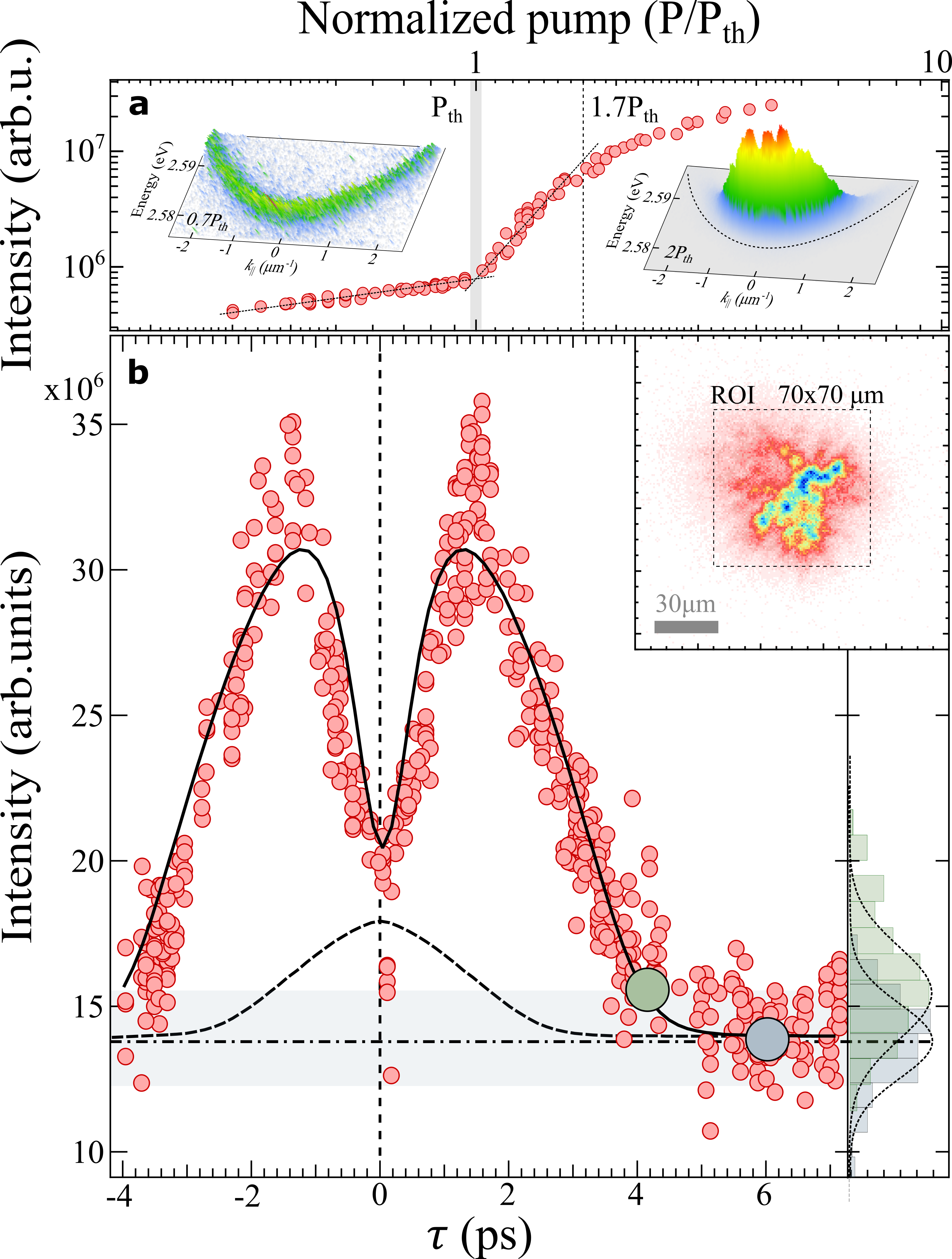}
    \vspace*{-0.25cm}
    \caption{\label{fig:2} \textbf{a} Polariton nonlinear emission intensity dependence vs the normalized pump fluence over $P_{\rm th}$. Colour-map insets show dispersion images of the lower polariton branch at below (left) and twice above (right) the polariton condensation threshold, $P_{\rm th}$ (see vertical, gray, and shaded line). \textbf{b} Single-shot, spatially integrated, polariton emission intensity from two consecutive, 200-fs pump-pulses vs pump-pump time-delay, $\tau$, at $P\cong 1.7P_{\rm th}$ for each pump-pulse. The dashed line corresponds to the k-independent theoretical model prediction, and the solid line depicts the best-fit result with the model, accounting for kinetic polariton-exciton bimolecular quenching. The right panel shows two normalized histograms of the emission intensity distribution comprised of $\sim100$ data points contained in a $\pm0.5$ ps time-interval around the plateau (blue dot) and the onset of amplification (green dot) regions, respectively. The top right inset shows a real-space, single-shot image of the polariton photoluminescence above $P_{\rm th}$ under single pump-pulse excitation, while the dashed-square highlights a $70\times70~\rm\upmu m$ region of interest (ROI) which was integrated to extract the single-pump-pulse $\rm\textbf{S}$-curve of panel (a) and the pump-pump intensity vs $\tau$ scan of panel (b).}
    \vspace*{-0.5cm}
\end{figure}

In this Letter, we develop a microscopic theoretical model accounting for non-resonant pumping, the cooling dynamics of the inhomogeneously broadened hot exciton reservoir, polariton thermalisation, vibron-mediated relaxation and kinetic polariton-exciton bimolecular quenching to investigate the intrinsic limitations of the temporal separation of consecutive polariton condensates. We have adapted the Lindbladian approach introduced in \cite{Zasedatelev2021} and further developed in \cite{Shishkov2022} that accounted for resonant excitation of polaritons in organic microcavities, and introduce to this model the polariton-exciton bimolecular quenching that occurs due to the delocalization of polaritons with large in-plane momenta. We apply our theoretical model to describe experimental observations of the temporal dynamics of polariton emission in the presence of two consecutive pump-pulses by varying their temporal separation and pulse-width. By combining real- and k-space analysis of the temporal dynamics of consecutive polariton condensates, we unravel the role of kinetic-induced bimolecular losses. Furthermore, by simulating the dynamics of polariton relaxation for a wide range of parameter-space we perform design optimization for future development of optical logic gates.

We begin with a microscopic model that describes the dynamics of the lower polariton branch (LPB) modes in a system with exciton-vibron optomechanical interactions, under pumping of a hot exciton reservoir tuned at one vibron above the ground state of the LPB to allow for resonant vibron-mediated energy relaxation. The governing Lindblad equation for the full density matrix $\hat{\rho}$ of the polaritons and vibrons is $\frac{d\hat{\rho}}{dt}=\frac{i}{\hbar}\left[\hat{\rho},\hat{H}_T\right]+\hat{L}_T(\hat{\rho})$. $\hat{H}_T$ is the Hamiltonian that includes the exciton-photon coupling and exciton-vibron optomechanical interactions. $\hat{L}_T(\hat{\rho})$ is the Lindblad-type superoperator describing radiative losses, cooling of the hot exciton reservoir, the pumping and thermalization processes which result from interactions of the LPB modes with the thermal bath of low energy molecular vibrations. By calculating the trace of multiplication of the density matrix with number-particle operators and by adiabatically excluding the vibron subsystem we arrive at the following system of coupled $N+1$ rate equations for the occupancy $n_i$ of $N$ LPB modes and the hot exciton reservoir, $n_{\rm hot}$:
\begin{equation}\label{eq:2}
   \vspace*{-0.25cm}
    \begin{gathered}
        \frac{dn_{\rm hot}}{dt}=-\gamma_{\rm hot} n_{\rm hot}+pump(t)-\\ -\sum_j G_j n_{\rm hot} (n_j+D_j)\\
        \frac{dn_i}{dt}=-\gamma_i n_i+ G_i n_{\rm hot} (n_i+D_i) +\\ +\sum_j\left[ \gamma_{\rm therm}^{j\rightarrow i} n_j (n_i+D_i)-\gamma_{\rm therm}^{i\rightarrow j} n_i (n_j+D_j)\right]
    \end{gathered}
\end{equation}
where $\gamma_{\rm hot} \equiv 1/ \tau_{\rm hot}$ is the cooling rate of non-resonantly injected hot excitons towards a dark, i.e. inactive exciton reservoir. The LPB decay rates, $\gamma_i$, account for the microcavity radiative losses and exciton decoherence. $G_i$ is the single-vibron transition rate stemming from exciton-vibron interactions, $\gamma_{\rm therm}$ is the inter-mode thermalization rate, and $D_i$ is the number of single particle modes in each of the $N$ rings in the k-space, appearing from the discretization of the model. However, for the parameters of the system that have been fixed in previous experiments \cite{Zasedatelev2021, Putintsev2024, Sannikov2024}, the model is almost k-independent within the small region of interest in the vicinity of $|\rm\textbf{k}|=0$, i.e. the thermalization processes are slow, resonant exciton-vibron scattering dominates, and $G_i$ and $\gamma_i$ depend weakly on $|\rm\textbf{k}|$, which make the two-mode approximation of the system valid (see Supplementary Information (SI), section I, for further information \cite{SI_references}). 

To investigate the intrinsic limitations of the temporal separation of consecutive polariton condensates, we utilise a $\pi-$conjugated ladder-type polymer, MeLPPP \cite{Schweitzer1999}, strongly coupled to the optical resonance of a conventional double distributed Bragg reflector planar microcavity. In this type of strongly coupled polymer microcavities, the hallmarks of polariton Bose-Einstein condensation were previously observed under non-resonant optical pumping \cite{Plumhof2013}, as well as vibron-mediated polariton relaxation \cite{Zasedatelev2019}. Here, we perform a single-shot imaging of the polariton emission both in real- and k-space and characterise the polariton condensate, under a single 200 fs pulse tuned at one vibron above the ground polariton state. Figure \ref{fig:2}a shows the dependence of the single-shot, polariton emission integrated over a $70\times70~\rm\upmu m$ region of interest (ROI) in real-space, see the top right inset of Fig.\ref{fig:2}b, on the incident fluence normalised on pump-threshold, $P_{\rm th}$ (gray vertical line), wherein we observe a clear onset of polariton condensation (see SI section II for full characterisation \cite{SI_references}). The two insets show characteristic dispersion images from below (left) and above (right) $P_{\rm th}$. Next, we introduce a consecutive pump-pump configuration with a variable time-delay between the pulses. We fix the incidence fluence of each pulse at $1.7P_{\rm th}$ and perform a time-delay scan. The recorded, single-shot, spatially-integrated, polariton emission is presented with red dots in Fig.\ref{fig:2}b. We observe two amplification maxima at time-delays ($\tau$) $\sim\pm1.7~\rm ps$. The non-monotonous $\tau$-dependence is phenomenologically attributed to the amplification of the residual polariton population that acts as a ``seed" for stimulating the relaxation of polaritons from the second pulse towards the LPB modes. The emission intensities at both zero- and large-time-delays ($|\tau|>5$ ps) are trivial and entirely governed by a single pulse experiment and correspond to the values of $I(2\times P)$ and $2\times I(P)$ intensity levels of the normalized pump power dependence of Fig.\ref{fig:2}a. For $|\tau|>2$ ps, with reducing the temporal separation between two consecutive pump pulses, the intensity of emission increases; see Fig.\ref{fig:2}b. This observation is in agreement with previous report, wherein polariton amplification into a ``seeded" mode exceeded in relaxation efficiency spontaneously-driven polariton condensation \cite{Zasedatelev2019}.

We apply our theoretical model to unravel the intrinsic dynamics of the competition between spontaneous and stimulated polariton relaxation, and describe the experimental observations of the temporal dynamics of polariton emission in the presence of two consecutive pump-pulses. The experimentally observed real-space polariton emission is proportional to the k- and time-integrated occupancies of LPB modes, $I(P)=\sum_{{\rm\textbf{k}}_i} \int_0^\infty \gamma_i n_i(t)dt$, that are the solutions of Eqs.\ref{eq:2}. We overlap the experimental data, presented with red dots in Fig.\ref{fig:2}b, with the numerical, k-space-integrated results (dashed line) and note that the discrepancy between theory and experiment originates from the k-independent polariron dynamics assumed in the theoretical model.

\begin{figure}[t!]
    \includegraphics[scale=1.15]{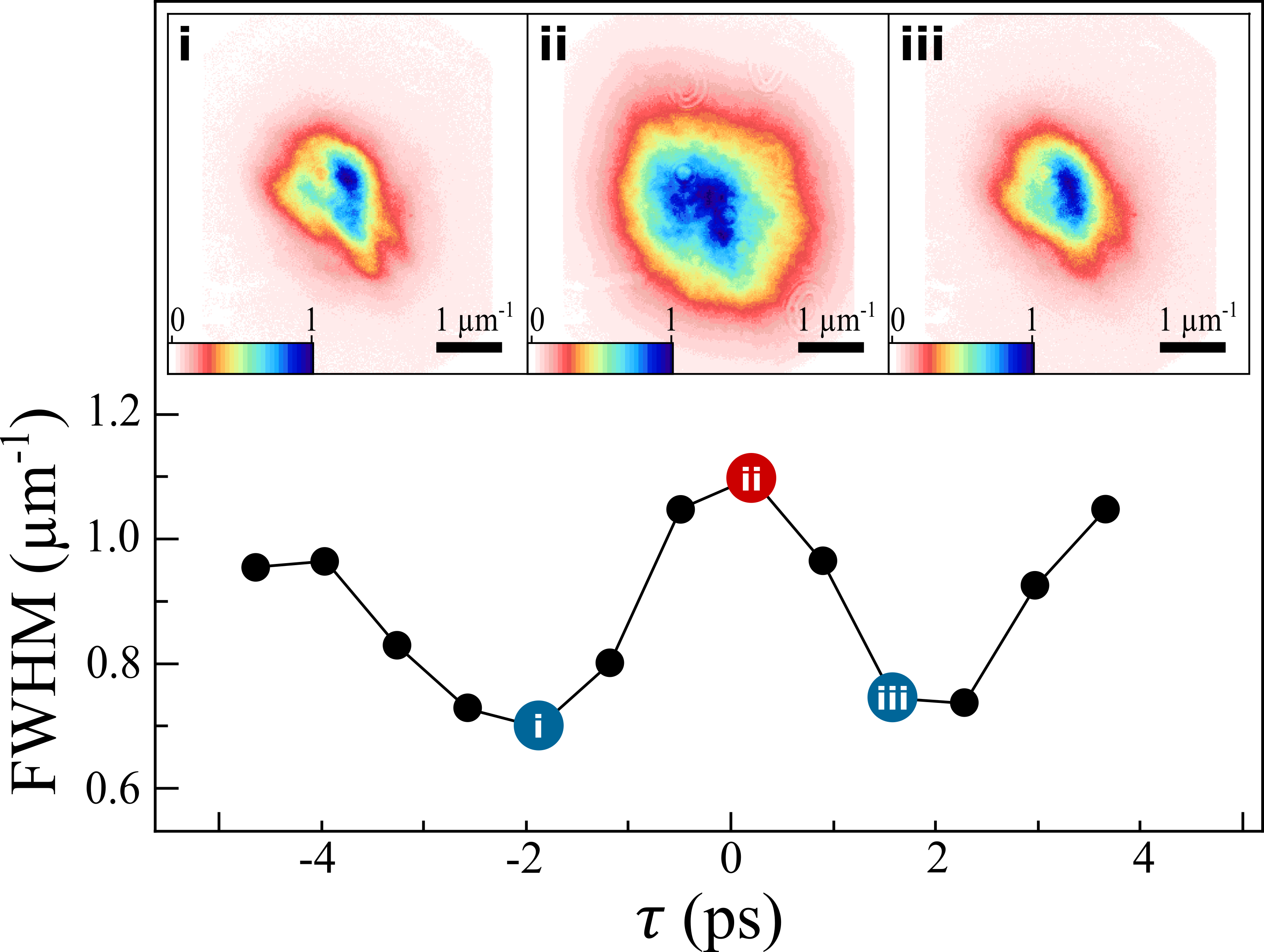}
    \vspace*{-0.25cm}
    \caption{\label{fig:3} Full width at half maximum (FWHM) of the experimentally measured, reciprocal-space, polariton emission profiles vs the time-delay, $\tau$, between consecutive pulses. Insets \textbf{i-iii} show the normalized, time-integrated, k-space images of the polariton emission for the respective points of the dependence.} 
    \vspace*{-0.5cm}
\end{figure}

To investigate the k-dependent polariton dynamics in the non-trivial time-delay dependence of Fig.\ref{fig:2}b, we perform time-integrated imaging of the k-space polariton emission for different pump-pump time-delays. Figure \ref{fig:3} shows the extracted full width at half maximum of the k-space emission profiles as a function of pump-pump time-delay, $\tau$. The insets of Fig.\ref{fig:3}, \textbf{i-iii}, correspond to three snapshots at time-delays annotated in Fig.\ref{fig:3}. We observe a clear narrowing of the distribution of polariton wavevectors towards the ground-state on either side of the zero time-delay, reaching a minimum at $\tau \sim\pm1.7~\rm ps$, that correlates with the emission maxima in Fig.\ref{fig:2}b. This wavevector redistribution points at the presence of k-dependent dynamics across the LPB that cannot stem from the slow, sub-nanosecond, multi-step thermalization processes \cite{Tereshchenkov2024}. The ultrafast wavevector redistribution is rather driven by rapid k-dependent non-radiative losses that gradually reduce the width of the distribution of the polariton wavevectors resulting in a relatively higher polariton occupancy towards \textbf{k}=0 upon arrival of the second 200 fs pump-pulse, see Fig.\ref{fig:3}. The physical mechanism behind such non-radiative losses of polaritons can be associated with bimolecular quenching of densely localized Frenkel excitons with the delocalized propagating polaritons of non-zero in-plane momenta \cite{Mazza2013, Zheng2024}. The additional non-radiative quenching channel increases linearly with the absolute value of polariton in-plane momentum: $\gamma_q({\rm\textbf{k}})=\gamma_q^{(1)}|\rm\textbf{k}|$, for derivation see SI, Section I. By taking into account such k-dependent losses we obtain a single-free-parameter model, which we use to fit the real-space time-delay scan on Fig.\ref{fig:2}b with the best fit value of $\gamma_q^{(1)}=4.8 ~\rm ps^{-1}\upmu m$, see the black solid line in Fig.\ref{fig:2}b. 

From our theoretical analysis, we find that for the motion induced non-radiative bimolecular quenching to influence the temporal dynamics of polariton condensation under consecutive pump-pulses, the temporal width of each pump-pulse should be shorter than the characteristic time of such process. For the best-fit parameter value that reproduces both the real- and k-space imaging of the pump-pump experiment, the kinetic losses are below one picosecond within the k-space ($|{\rm\textbf{k}}|<3~\rm\upmu m^{-1}$). To confirm the validity of our model in describing the limitations of our system on the temporal bandwidth of consecutive polariton condensates, we perform a control experiment, where we use 2.5 ps pump-pulses. Under such pumping conditions, we do not observe any evidence of a time-delayed polariton amplification, see SI section III \cite{SI_references}.

By comparing the time-delay scan under 2.5 ps and 200 fs pump-pulses, we find that the half width at half maximum of the intensity profile changes from $\sim 2.6~\rm ps$ to $\sim 2.4~\rm ps$ for the shorter pump-pulses, predominantly due to the presence of motion-induced polariton losses and the concomitant emergence of time-delayed polariton amplification. For the shorter pulses used here, we find that the minimum temporal separation of consecutive polariton condensates is limited to $4.2~\rm ps$, defined from the analogue of the Rayleigh criterion for the determination of the plateau region of the time-delay scan at longer times, see histograms on the right panel of Fig.\ref{fig:2}b. Whereas the maximum operation bandwidth is limited to 240 GHz, we note that the one parameter that dominates this limit, and does not depend on the intrinsic photophysics of the polymer, is the polariton lifetime. By varying both the exciton-photon detuning and the $Q$-factor of the cavity, we find that the maximum operation bandwidth may be substantially increased approaching 0.5 THz (see SI, section IV for device optimisation simulations \cite{SI_references}). 

%CONCLUSION
In conclusion, we investigate the intrinsic processes limiting the temporal separation of consecutive polariton condensates in a polymer microcavity. For pump-pulse temporal widths shorter than any characteristic time of the system, we find that beyond the radiative polariton lifetime, kinetic bimolecular quenching of polaritons dominates the dynamics. We have developed a microscopic model that describes the dynamics of the system under consecutive pump-pulses. Through device optimisation we predict sub-THz bandwidth as the upper limit for the operation of polariton transistors, signifying the importance for future progress in room-temperature polaritonics towards ultrafast all-optical logic.
\\
\\

The authors sincerely thank D. Urbonas, T. St\"oferle, and R. F. Mahrt for useful discussions and the provision of microcavity samples. This work was supported by the Russian Scientific Foundation (RSF) Grant No. 23-72-00059 to D.S..

\setcounter{equation}{0}
\setcounter{figure}{0}
\setcounter{section}{0}
\setcounter{subsection}{0}
\newcolumntype{P}[1]{>{\centering\arraybackslash}p{#1}}
\newcolumntype{M}[1]{>{\centering\arraybackslash}m{#1}}
\renewcommand{\theequation}{S\arabic{equation}}
\renewcommand{\thefigure}{S\arabic{figure}}
\renewcommand{\baselinestretch}{1}
\onecolumngrid
\newpage
\vspace{1cm}
\begin{center}
\Large \textbf{Supplementary Information}
\end{center}

\section{Section I}

In this section, we describe the microscopic model used to simulate system dynamics and present the full best-fit parameter set. We also vary the model parameters in the reasonable range to investigate the sensitivity of maximum repetition rate and amplification of the system. We provide the reasoning for the introduction of kinetic losses in the system and explain the corresponding derivations.

\subsection{Differential equations and discretization}

Stemming from the microscopic model described in \cite{Zasedatelev2021, Shishkov2022}, we modify the system of coupled equations for the occupancies of the lower polariton branch (LPB) modes by introducing the non-resonant pump, ``injecting" excitons in a quasi-state at one vibron above the ground state of the LPB. In such configuration no resonant transitions are possible into the upper polariton branch, therefore we neglect the corresponding equations from \cite{Zasedatelev2021}. We also note that the hot exciton reservoir is being rapidly cooled \cite{Dai2013} through non-radiative channels, unlike the pumped upper polariton mode in the resonant pump configuration. We obtain the following continuous set of coupled equations in the limit of fast vibron decay for the number of LPB polaritons with wavevector ${\rm\textbf{k}}$, $n_{\rm\textbf{k}}$, and the equation for hot exciton reservoir occupancy, $n_{\rm hot}$:
\begin{equation}\label{eq:s1}
    \begin{gathered}
        \frac{dn_{\rm hot}}{dt}=-\gamma_{\rm hot} n_{\rm hot}+pump(t)-\sum_{\rm\textbf{k}} G_{\rm\textbf{k}}^{\rm hot} n_{\rm hot} (n_{\rm\textbf{k}}+1)\\
        \frac{dn_{\rm\textbf{k}}}{dt}=-\gamma_{\rm\textbf{k}} n_{\rm\textbf{k}}+ G_{\rm\textbf{k}}^{\rm hot} n_{\rm hot} (n_{\rm\textbf{k}}+1) +\\ +\sum_{\rm\textbf{k}}\left[ \gamma_{\rm therm}^{{\rm\textbf{k'}} \rightarrow {\rm\textbf{k}}} n_{\rm\textbf{k'}} (n_{\rm\textbf{k}}+1)-\gamma_{\rm therm}^{{\rm\textbf{k}}\rightarrow {\rm\textbf{k'}}} n_{\rm\textbf{k}} (n_{\rm\textbf{k'}}+1)\right]
    \end{gathered}
\end{equation}
Here, $\gamma_{\rm hot}$ is the cooling rate of hot excitons, and $\gamma_{\rm\textbf{k}}$ are the radiative losses of polaritons associated with the cavity lifetime and exciton decoherence. We estimate polariton losses as follows:

\begin{equation}\label{eq:s2}
    \begin{gathered}
        \gamma({\rm\textbf{k}})=\gamma_{\rm cav}|C_{\rm phot}({\rm\textbf{k}})|^2+\gamma_{\rm exc}|C_{\rm exc}({\rm\textbf{k}})|^2 \approx \gamma_{\rm g.s.}+\gamma_2 {\rm\textbf{k}}^2\\
        \gamma_2=(\gamma_{\rm exc}-\gamma_{\rm cav})\frac{\alpha \Omega_R^2}{4(\Omega_R^2+\Delta^2/4)^{3/2}}
    \end{gathered}
\end{equation}
The estimated value of $\gamma_2$ for our system parameters is of the order of $\sim0.1~\rm ps^{-1}\upmu m^2$ which is insignificant considering the condensate FWHM in the momentum space is $\sim1~\rm\upmu m^{-1}$. We notice that resonant single-vibron transition rates from the reservoir $G_{\rm\textbf{k}}^{\rm hot}$ are obtained through adiabatic exclusion of the vibron subsystem and are therefore of Lorenzian shape:
\begin{equation}\label{eq:s3}
    G_{\rm\textbf{k}}^{\rm hot}=\frac{g^2 X_{\rm\textbf{k}} (\gamma_{\rm hot}+\gamma_{\rm\textbf{k}}+\gamma_{\rm vib})}{(\omega_{\rm hot}-\omega_{\rm\textbf{k}}-\omega_{\rm vib})^2+\frac{1}{4}(\gamma_{\rm hot}+\gamma_{\rm\textbf{k}}+\gamma_{\rm vib})^2}
\end{equation}
where $g$ is the exciton-vibron interaction constant, $X_{\rm\textbf{k}}$ is the exciton fraction of the corresponding LPB mode. $\omega_{\rm vib}$ and $\gamma_{\rm vib}$ are the frequency and spectral width of the main molecular vibron quantum of MeLPPP polymer with $\nu=2$ (see Ref.\cite{Zasedatelev2019}). We note however that such approach only accounts for the case of single narrow-linewidth hot exciton mode, which is not the case for MeLPPP. Therefore, in order to achieve correct effective transition rate we need to average the expression \ref{eq:s3} over $\omega_{\rm hot}$ in the vicinity of the central pump wavelength $\langle \omega_{\rm hot} \rangle = 2.8$ eV with the exciton linewidth being equal to the experimentally measured inhomogeneous broadening $\sigma_{\omega_{\rm hot}} \approx 60$ meV. Since the LPB region of interest in this study is $|{\rm\textbf{k}}|<3~\rm\upmu m^{-1}$ (from experimental observations the condensate only forms within $|{\rm\textbf{k}}|<1~\rm\upmu m^{-1}$), the width of this region, $\Delta = \alpha_{\rm cav}k_{\rm max}^2=19.8$ meV is at least 4 times smaller than width of the resonance. This gives us reasonable grounds to assume that for each LPB mode within the region of interest the rate of transitions is very close and small variations of $G$ depend more on the inhomogeneity of the reservoir than on deterministic mode parameters. We assume $G$ to be k-independent, with the average value of $\langle G\rangle\approx10^{-7}$ eV for $g=0.1$ meV. We also note that $G$ quadratically depends on $g$, and $g$ is taken as a model parameter, whose value is reasonably estimated to be between 0.1 and 0.5 meV \cite{Zasedatelev2021}. Therefore, we can carefully vary $G$ in agreement with previous results.\\

Finally, $\gamma_{\rm therm}^{{\rm\textbf{k}}\rightarrow {\rm\textbf{k'}}}$ is the intermode thermalization rate
originating from the interactions of the LPB modes with a bath of lower energy molecular vibrations. The downwards processes ($\omega_{\rm\textbf{k}}>\omega_{\rm\textbf{k'}}$) are considered to be frequency-independent, $\gamma_{\rm therm}^{ {\rm\textbf{k}}\rightarrow{\rm\textbf{k'}}}=\gamma_{\rm therm}$, whereas the inverse processes obey the Kubo-Martin-Schwinger relation: $\gamma_{\rm therm}^{ {\rm\textbf{k}}\rightarrow  {\rm\textbf{k'}}}=\gamma_{\rm therm}\exp{(-\Delta \omega /T)}$ for $\omega_{\rm\textbf{k}}<\omega_{ {\rm\textbf{k'}}}$. The value of $\gamma_{\rm therm}$ is taken as $3*10^{-9}$ eV as the best fit result from previous numerical simulations, achieving great agreement with an even more complex experimental configuration \cite{Sannikov2024}. We notice that the later theoretical estimation of thermalization predicted an even smaller value, below $10^{-9}$eV. Since the thermalization rate is much smaller than the resonant transition rate $G$, the modes interact with each other indirectly, through reservoir - any small preoccupation of polaritons in one mode results in ultra-fast draining of the reservoir into it, not leaving many excitons to scatter into other modes.\\

The system is then discretized. Due to cylindrical geometry of the experiment we ``slice" the LPB dispersion paraboloid for $|{\rm\textbf{k}}|<k_{\rm max}=3~\rm\upmu m^{-1}$ into rings, equidistant in $\omega$ domain. Since the system is two dimensional and the dispersion is parabolic, we obtain uniform density of states, i.e. there are equal numbers of single particle modes in each of the rings, and, therefore, their contribution to the time and k-space integrated emission, recorded in the experiment (see main text), is the same. As a result, we rewrite the system of Eqs.\ref{eq:s1} for N discrete rings: 

\begin{equation}\label{eq:s4}
    \begin{gathered}
        \frac{dn_{\rm hot}}{dt}=-\gamma_{\rm hot} n_{\rm hot}+pump(t)-\sum_j G_j n_{\rm hot} (n_j+D_j)\\
        \frac{dn_i}{dt}=-\gamma_i n_i+ G_i n_{\rm hot} (n_i+D_i) +\\ +\sum_j\left[ \gamma_{\rm therm}^{j\rightarrow i} n_j (n_i+D_i)-\gamma_{\rm therm}^{i\rightarrow j} n_i (n_j+D_j)\right]
    \end{gathered}
\end{equation}

With thermalization being slow, $G$ and $\gamma_i$ independent on k the model can be considered k-independent, allowing for a two-equation approximation with $n_p=\sum_i n_i$, the total number of single particle states $D_p=\sum_i D_i$, and $\gamma_p=\gamma_i$:
\begin{equation}\label{eq:s5}
    \begin{gathered}
        \frac{dn_{\rm hot}}{dt}=-\gamma_{\rm hot} n_{\rm hot}+pump(t)-G n_{\rm hot}(n_{p}+D_{p})\\
        \frac{dn_p}{dt}=-\gamma_p n_p +G n_{\rm hot}(n_{p}+D_{p})
    \end{gathered}
\end{equation}

The recorded real- and k-space non-resolved polariton emission is then calculated as $I=\int_0^\infty \gamma_p n_p(t)dt$. Though such model does describe the nonlinear dependence of $I$ on the pump power, the effect of strong amplification of filtered-in-k-space polariton emission when the selected mode is resonantly seeded with few polaritons at the expense of other modes can not be described \cite{Zasedatelev2019,Zasedatelev2021,Sannikov2024}. The real space pump-pump time-delay scan, presented in this work cannot be described through such model as well - see dashed line on Fig.\ref{fig:2}b of the main text.
\newpage
\subsection{Introduction of kinetic bimolecular quenching}

Since the total amount of hot excitons injected into the system does not depend on a time-delay between the two pump pulses, and equals $2\times P$, there should be some non-radiative channel that is responsible for the non-trivial behaviour of the real-space emission shown in Fig.\ref{fig:2}b of the main text. We notice that the cooling of hot excitons, $\gamma_{\rm hot}$, which is a non-radiative loss channel of hot excitons, alone cannot account for this effect, since in the above-threshold regime more than 80\% of hot excitons are already condensed into the LPB, capping the potential amplification at $\sim 25\%$, which is smaller than the observed amplification. From the correlation between the amplification of the real-space emission and the narrowing of the k-space (Fig.\ref{fig:2}b and Fig.\ref{fig:3} of main text, respectively), we deduce that the effect of such non-radiative channel should be stronger for condensates with wider k-vector distribution. This hints at the presence of k-dependent non-radiative channel of polariton losses.\\  
The loss of energy in a strongly excited semiconductor, i.e. with large exciton density, can be attributed to the bimolecular quenching of excitons, that happens non-radiatively. However, it is well known that Frenkel excitons in organic polymers cannot participate in such process due to strong localization and absence of wavefunction overlap. The polaritons, however, are delocalized, and the area of the delocalization grows proportionally to the square of the in-plane momentum absolute value, $A_{\rm deloc} \propto |{\rm\textbf{k}}|^2$. For the high density of excitons this would mean that the chance for a polariton to encounter the exciton in its lifetime is proportional to $|{\rm\textbf{k}}|^2$. The probability of quenching however is proportional to the exciton and polariton wavefunction overlap: $p \propto \int \psi_x^*\psi_{p}d{\rm\textbf{r}}$. Due to the normalization of a polariton wavefunction, the amplitude of $\psi_p$ is proportional to $1/k$. The effective quenching losses for polaritons with in-plane momentum k are then:
\begin{equation}\label{eq:s6}
    \gamma_q({\rm\textbf{k}}) \propto A_{\rm deloc} \int \psi_x^*\psi_{p}d{\rm\textbf{r}} \propto |{\rm\textbf{k}}|
\end{equation}
We, therefore, introduce these losses into Eqs.\ref{eq:s3} with some arbitrary proportionality coefficient: $\gamma_q({\rm\textbf{k}})=\gamma_q^{(1)}|{\rm\textbf{k}}|$. We arrive at the system of equations with just one free parameter, $\gamma_q^{(1)}$:
\begin{equation}\label{eq:s7}
    \begin{gathered}
        \frac{dn_{\rm hot}}{dt}=-\gamma_{\rm hot} n_{\rm hot}+pump(t)-\sum_i G n_{\rm hot}(n_{i}+D_{i})\\
        \frac{dn_i}{dt}=-(\gamma_i+\gamma_q(k_i))n_i + G n_{\rm hot}(n_{i}+D_{i})
    \end{gathered}
\end{equation}
The calculated emission from each mode is then: $I_i=\int_0^\infty \gamma_i n_i(t) dt$, and the real-space emission that we compare with the experimentally measured light emission from the sample is $I_{\rm RS}=\sum_i I_i$. The introduced mechanism of k-dependent losses narrows the remaining polariton population in time after the formation of the first condensate. It results in a more effective draining of the hot exciton reservoir (replenished by the second pulse) into the ground state, which is least exposed to non-radiative energy loss. This results in the formation of a peak in real-space emission at $\tau\sim\pm1.7~\rm ps$

\newpage

\subsection{Fit of the experimental dependencies}
Here, we perform numerical simulations of Eqs.\ref{eq:s7} with fixed from the preevious simulations $G$, $\gamma_i$, $D_i$ while varying the free parameter $\gamma_q^{(1)}$ In Table \ref{tab:1} we provide the best-fit parameter set and their justifications, as well as additional parameters, used to calculate $G$, cavity dispersion and lifetimes.

\begin{table}[h!]\label{tab:1}
    \centering
    \begin{tabular}{ |M{3.5cm}|M{2cm}|M{5cm} |}
         \hline
    %\multicolumn{2}{|c|}{\textbf{Simulation parameters}} \\
    \textbf{Parameter} & \textbf{Value} & \textbf{Grounds}\\
    
    \hline
     \(\bm{\omega_{\rm cav}}~\rm{[eV]}\) & 2.65 & Fit of k-space image of the system below threshold (Fig.\ref{SI:1}b)  \\
     \hline
     \(\bm{\omega_{\rm exc}}~\rm{[eV]}\) & 2.72 & Fit of the absorption spectrum of MeLPPP \cite{Plumhof2013}  \\ 
     \hline
     \(\bm{\gamma_{\rm cav}}~\rm{[meV]}\) & 2.2 &  Fit of k-space image of the system below threshold (Fig.\ref{SI:1}b)  \\
     \hline
     \(\bm{\gamma_{\rm hot}}~\rm{[ps^{-1}]}\) & 1 & Cooling time of hot excitons ~1ps \cite{Dai2013}\\
     \hline
     \(\bm{\gamma_{\rm exc}}~\rm{[meV]}\) & 50 & Corresponds to decoherence of excitons \cite{Plumhof2013}\\
     \hline
     \(\bm{\alpha_{\rm cav}}~\rm{[meV\cdot\upmu m^2]}\) & 2.2 & Fit of k-space image of the system below threshold (Fig.\ref{SI:1}b)\\
     \hline
     \(\bm{\Omega_{R}}~\rm{[meV]}\) & 85 & Fit of k-space image of the system below threshold (Fig.\ref{SI:1}b)\\
     \hline
     \(\bm{\omega_{\rm hot}}~\rm{[eV]}\) & 2.8 & The laser is tuned one vibron quantum above the ground state\\
     \hline
     \(\bm{\omega_{\rm vib}}~\rm{[meV]}\) & 200 & Fit of the absorption spectrum \cite{Zasedatelev2021}\\
     \hline
     \(\bm{\gamma_{\rm vib}}~\rm{[meV]}\) & 2.5 & Fit of the Raman spectrum \cite{Zasedatelev2021}\\
     \hline
     \(\bm{g}~\rm{[meV]}\) & 0.1 & Best fit parameter from previous experiments, agrees well with theoretical approach in \cite{Shishkov2019}\\
     \hline
     \(\bm{T}~\rm{[meV]}\) & 25 & Experiments were conducted at room temperature\\
     \hline
     \(\bm{\gamma_{\rm therm}}~\rm{[eV]}\) & $3\cdot10^{-9}$ & Best fit model parameter from previous experiments with the sample 
     \cite{Zasedatelev2021,Sannikov2024}\\
     \hline
     \(\bm{\gamma_q^{(1)}}~\rm{[ps^{-1}\cdot \upmu m]}\) & $4.8$ & Quenching coefficient, free parameter\\
     \hline
     \(\bm{G}~\rm{[ps^{-1}]}\) & $10^{-4}$ & Calculated from parameters above\\
     \hline
     \(\bm{\gamma_i}~\rm{[ps^{-1}]}\) & $3.33$ & Calculated from parameters above\\
     \hline
    \end{tabular}
    \caption{The parameters of the cavity under study used for numerical simulations.}
    \label{tab:1}
\end{table}
\newpage

\subsection{Estimation of amplification maxima positions}

In this section, we evaluate the dynamic narrowing of the polariton condensate before the arrival of the second pulse under the action of kinetic quenching. We first consider the dynamics of the polariton distribution function $f({\rm\textbf{k}},t)$ (henceforth $|{\rm\textbf{k}}|\equiv k$) after the condensate is formed without any additional excitations. We investigate the dynamics within the few picosecond timescale, therefore the re-distribution of polaritons across the modes due to thermalization can be omitted. The deviation from the initial polariton distribution $f_0(k)$ is only driven by the k-dependent losses:

\begin{figure}[b!]
    \centering
    \includegraphics[scale=1]{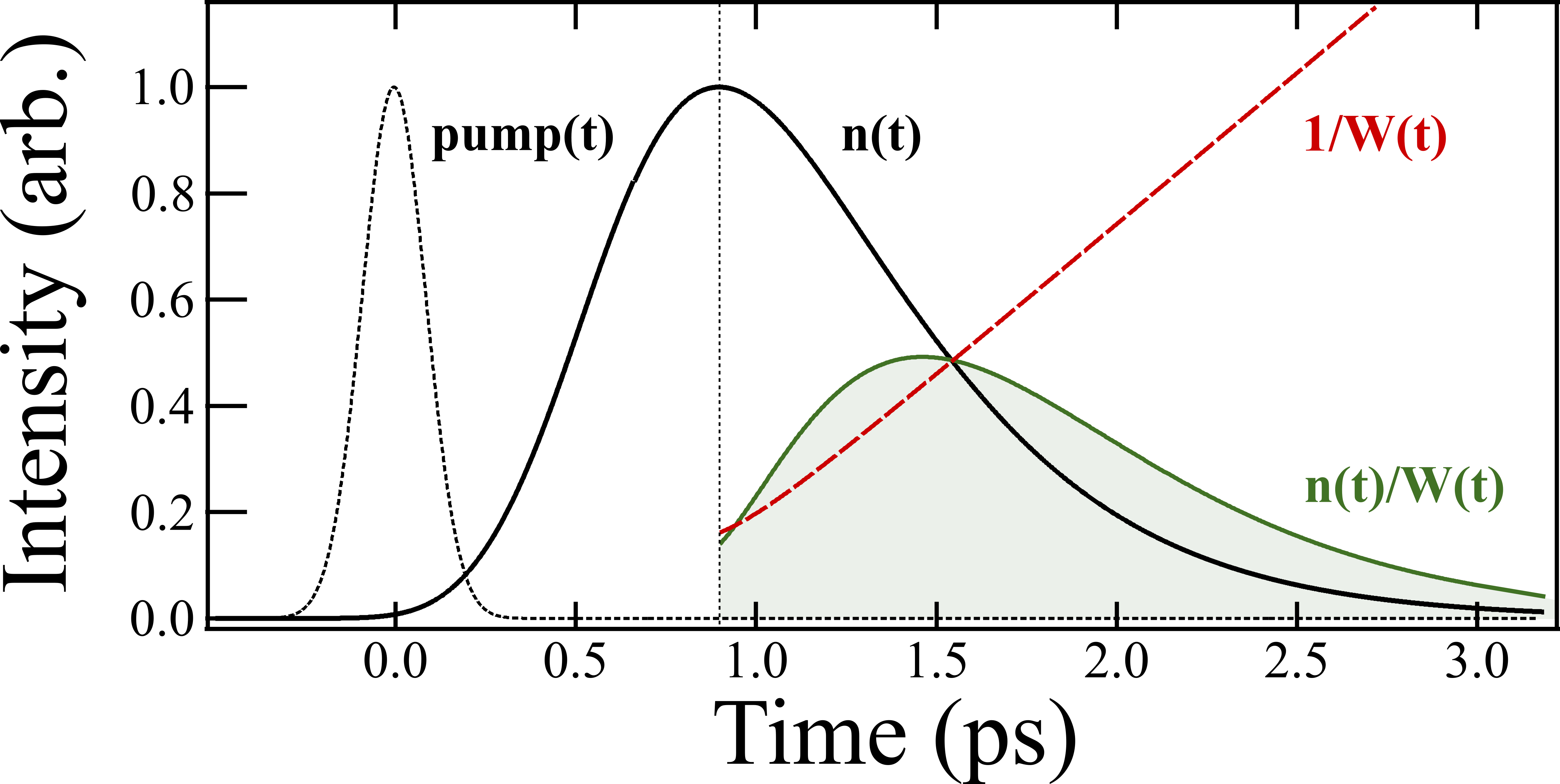}
    %\vspace*{-0.25cm}
    \caption{\label{SI:1} Time-resolved dynamics of polariton condensate. The normalized pumping pulse (black-dashed line), the polariton occupancy of the ground state (black-solid line), and the inverse full width at half maximum of the polariton condensate wavevector distribution (red-dashed line) as a function of time. The expected amplification efficacy (green-solid line) has a maximum around $\sim1.5~\rm ps$ defining an approximate moment in time, $\tau$, for sending the second pump pulse to observe the maximum polariton amplification.} 
    %\vspace*{-0.5cm}
\end{figure}

\begin{equation}\label{eq:s8}
    f(k,t) = f_0(k)e^{-\gamma_0 t- \gamma_q^{(1)} k t}
\end{equation}
From single-pulse experimental observations it is reasonable to assume the initial distribution to be gaussian, $f_0(k)=C \exp{(-k^2/\sigma^2)}$. We estimate $\sigma$ from the experimentally measured time-integrated polariton emission, 
\begin{equation}\label{eq:s9}
    I(k)\sim \gamma_0 \int_0^\infty \exp{(-k^2/\sigma^2-\gamma_0 t- \gamma_q^{(1)} k t)} dt=\frac{\exp(-k^2/\sigma^2)}{1+\frac{\gamma_q^{(1)} k}{\gamma_0}}
\end{equation}
The experimental FWHM for this distribution is known, $W_e=1.1~\rm\upmu m^{-1}$, which together with Eq.\ref{eq:s9} gives us $\sigma=0.5 W_e/\sqrt{-\ln \left(0.5\left(1+\frac{\gamma_q^{(1)} W_e}{2\gamma_0}\right)\right)}\approx 0.89~\rm\upmu m^{-1}$. We than estimate how the FWHM of the condensate in the k-space evolves, $W=2k_*$, from the equation:
\begin{equation}\label{eq:s5}
    \frac{1}{2}=\frac{f(k_*,t)}{f(0,t)}=\exp(-(k_*/\sigma)^2-\gamma_q^{(1)} k_* t)
\end{equation}
After introducing the dimensionless units $x_*=k_*/\sigma$ and $\beta = \gamma_q^{(1)} \sigma t/2$ we obtain a quadratic equation for $x_*$:
\begin{equation}
    x_*^2 +2\beta x_* -\ln 2 =0
\end{equation}
which gives the desired FWHM dynamics:
\begin{equation}
    W(t)=2\sigma (-\beta +\sqrt{\beta^2+\ln 2})
\end{equation}

The amplification can roughly be estimated as proportional to both the remaining polariton population and the inverse of the $W(t)$. We plot this product together with the pumping pulse and polariton population obtained from the numerical simulations in Fig.\ref{SI:1}, and observe a clear picosecond dynamics, with the amplification peak at $\sim 1.5~\rm ps$, which corresponds well with the experimentally observed value of $1.7~\rm ps$.

\subsection{Numerical simulations of k-resolved time-delay intensity measurements}

\begin{figure}[h!]
    \centering
    \includegraphics[scale=1.3]{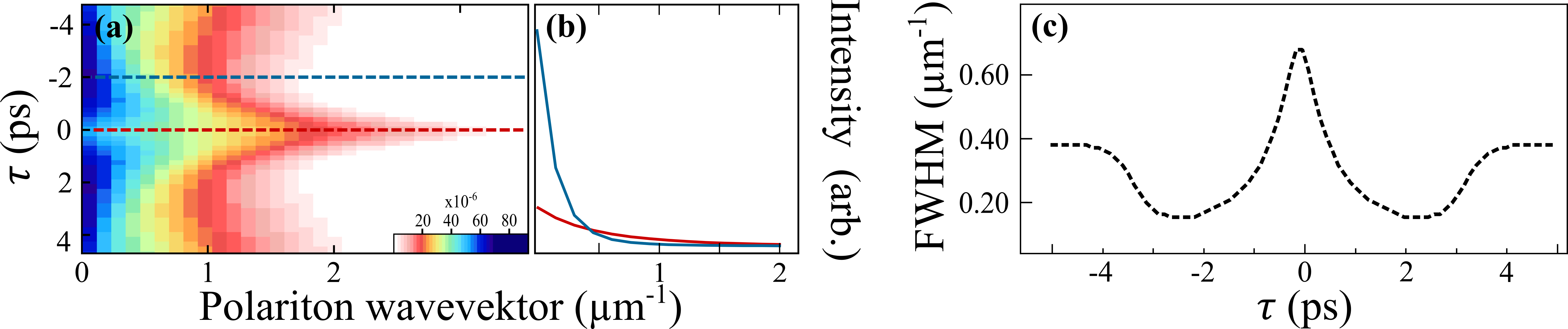}
    %\vspace*{-0.25cm}
    \caption{\label{SI:2} \textbf{(a)} Colormap of polariton emission intensity vs time-delay, $\tau$, between consecutive pump pulses (vertical axis) and polariton wavevector (horizontal axis). \textbf{(b)} show the polariton density/emission intensity extracted from the cross-section of the dashed lines of the colour-map of \textbf{a} at zero time-delay (red lines) and the time-delay corresponding to the maximum of the amplification (blue lines), respectively. \textbf{(c)} Full width at half maximum (FWHM) of the modeled polariton emission profiles in reciprocal-space vs $\tau$.} 
    %\vspace*{-0.5cm}
\end{figure}

We perform the numerical simulations of Eqs.\ref{eq:s7} for a range of time-delays, $\tau$, and plot the recorded emission intensity for each value of the in-plane momentum in the form of a colormap in Fig.\ref{SI:2}a. It is visible that indeed the amplification happens in the least lossy modes, in the vicinity of the ground state $k=0$. To study the change in the time-integrated polariton distribution shape for different time-delay values we plot the two slices of the colormap at $\tau=0 ~\rm ps$ and $\tau=1.7~\rm ps$ with red and blue lines respectively on panel Fig.\ref{SI:2}b. For the latter case, the excitons injected with the second pulse have scattered predominantly into $k=0$ mode unlike the case of a single pulse spontaneous polariton condensation (zero time-delay, red curve). We extract the characteristic width of such polariton distribution for each value of $\tau$, and plot it in Fig.\ref{SI:2}c, reporducing the main features of the experimentally extracted width of polariton emission in k-space (see Fig.\ref{fig:3} of the main text). 

\newpage

\section{Section II}

\begin{figure}[h!]
    \centering
    \includegraphics[scale=0.82]{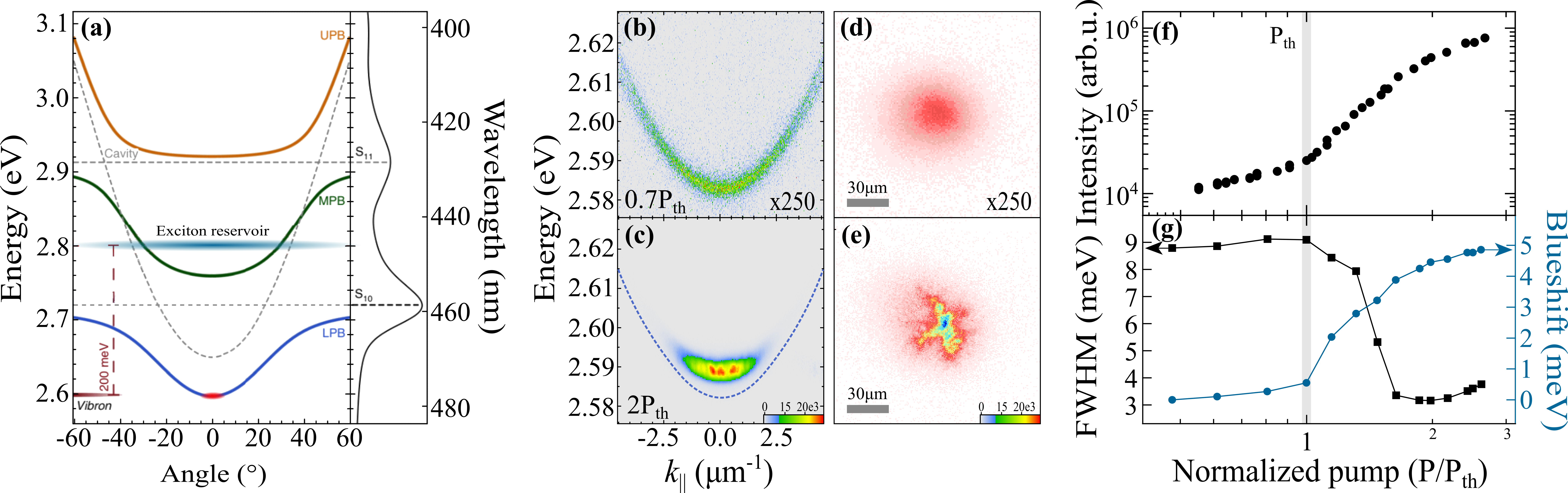}
    %\vspace*{-0.25cm}
    \caption{\label{SI:3} Hallmarks of the Bose-Einstein condensation. \textbf{(a)} Schematic of upper/middle/lower polariton dispersions drawn in orange, green, and blue solid lines, respectively. Dashed lines indicate the bare cavity mode (Cavity) and the two sub-levels of the first excited singlet state of MeLPPP (S$_{10}$ and S$_{11}$). The optical pump tuned at 2.8 eV photon energy populates exciton reservoir depicted with a blue-shaded horizontal stripe. The vibron-mediated exciton-to-polariton relaxation process depicted by the vertical red dashed-line. Time-integrated dispersion images of the polariton photoluminescence from the lower polariton branch at \textbf{(b)} below and \textbf{(c)} above the condensation threshold. Real-space, single-shot images of the polariton photoluminescence \textbf{(d)} below and \textbf{(e)} above $P_{\rm th}$ under single pump-pulse excitation. Dependencies of the \textbf{(f)} real-space polariton emission intensity and \textbf{(g)} the ground-state polariton emission linewidth (black squares, left axis) and blueshift (blue circles, right axis) as a function of pump fluence normalized at condensation threshold.} 
    %\vspace*{-0.5cm}
\end{figure}

In this Section, we show hallmarks of a Bose-Einstein condensation in a strongly-coupled microcavity filled with a $\pi-$conjugated ladder-type polymer MeLPPP. The structure under study consists of a 35-nm-thick neat film of MeLPPP between 50-nm SiO$_2$ spacers, sandwiched between SiO$_2$/Ta$_2$O$_5$ distributed Bragg reflectors on a glass substrate. Strong coupling of the cavity mode (2.65 eV) and two sub-levels of the first excited singlet state (S$_{10}$ at 2.72 eV and S$_{11}$ at 2.91 eV, as shown in the absorption spectrum plotted on the right panel of Fig.\ref{SI:3}a, of MeLPPP results in three polariton branches exhibiting 144 meV energy splitting between the middle and the lower polariton branches (Rabi splitting), which is twice as large as the exciton–photon detuning (-70 meV). Figure \ref{SI:3}a shows a schematic of upper/middle/lower polariton dispersion relations (orange, green, and blue solid lines) with corresponding bare cavity mode and first excited singlet state of MeLPPP (gray dashed lines), and the optical excitation that feeds a hot exciton reservoir of the polariton system (blue shaded horizontal stripe). The optical pumping is tuned at 2.785 eV one molecular vibron ($\approx200~\rm{meV}$) above the lower polariton branch at 2.585 eV. Time-integrated energy-momentum distributions of the polariton photoluminescence from the lower polariton branch at below and twice above the condensation threshold are shown in Fig.\ref{SI:3}b,c, respectively. Single-shot real-space polariton emission at below and twice above the condensation threshold are shown in Fig.\ref{SI:3}d,e, respectively. Analysis of the single-shot real-space emission clearly evidences superlinear increase of a polariton occupancy above the threshold of $\sim40~\rm\upmu J cm^{-2}$ of absorbed incident pumping fluence in Fig.\ref{SI:3}f, while the spectrally resolved polariton emission integrated over \textit{k}$_{\parallel}$ = 0 $\pm1~\rm\upmu m^{-1}$ in the reciprocal space manifests a significant line narrowing and blueshift, see Fig.\ref{SI:3}g. These features are commonly considered as hallmarks of polariton condensation in organic microcavities.

\section{Section III}
\begin{figure}[h!]
    \centering
    \includegraphics[scale=1]{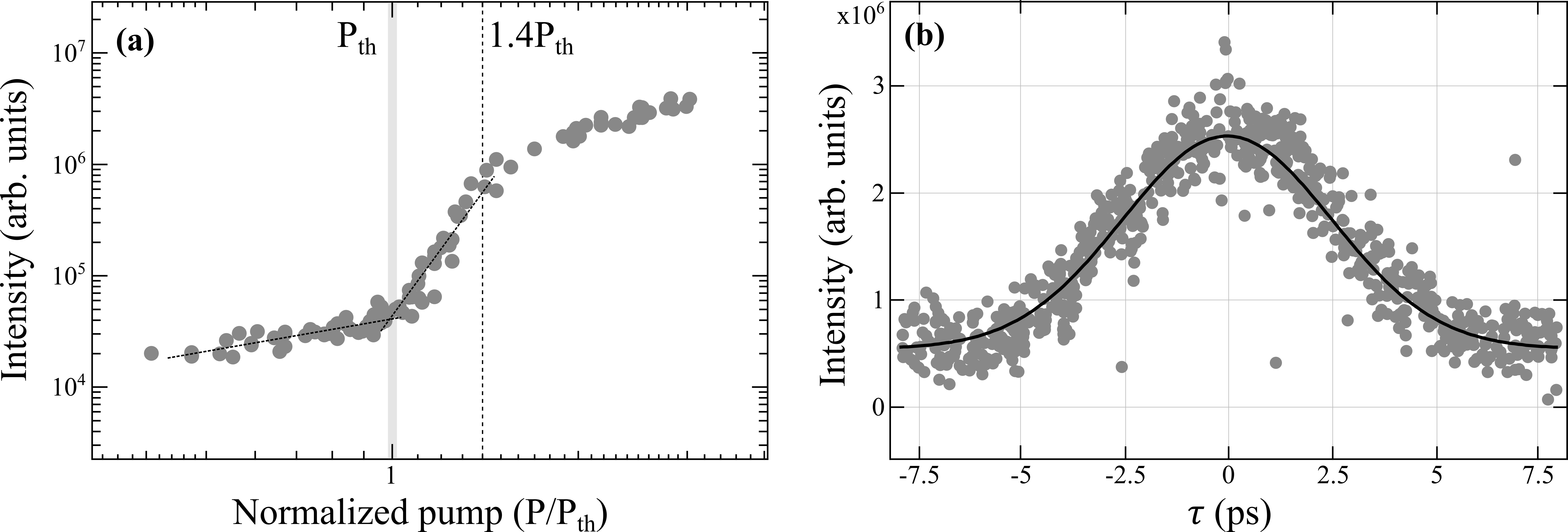}
    %\vspace*{-0.25cm}
    \caption{\label{SI:4} \textbf{(a)} Real-space, single-shot polariton emission intensity dependence vs the normalized pump fluence over $P_{\rm th}$ under a single-, 2.5-ps-pumping pulse excitation. \textbf{(b)} The output polariton emission intensity from two adjacent 2.5-ps-pumping pulses as a function of a time-delay between them, each set at $P\cong 1.4P_{\rm th}$ of normalized incident fluence with the black-solid line representing a model fit. } 
   \vspace*{-0.5cm}
\end{figure}

In this Section, we perform system analysis for longer pumping pulses to confirm the origin of the k-space redistribution. We elongate the pumping pulses up to 2.5 ps setting each of them at $\approx1.4P_{\rm th}$ above the condensation threshold, see the intensity  $\rm\textbf{S}$-curve and the vertical dashed line in Fig.\ref{SI:4}a, and repeat the time-delay scan. In such regime the k-dependent polariton losses do not allow for effective mode competition within LPB due to a long quasi-continuous feeding of these modes from the hot exciton reservoir. Therefore, neither the polariton amplification nor the k-space redistribution are observed. We plot the corresponding experimental (gray dots) and theoretical (black line) in Fig.\ref{SI:4}b.

\section{Section IV}

 \begin{figure}[b!]
    \centering
    \includegraphics[scale=0.9]{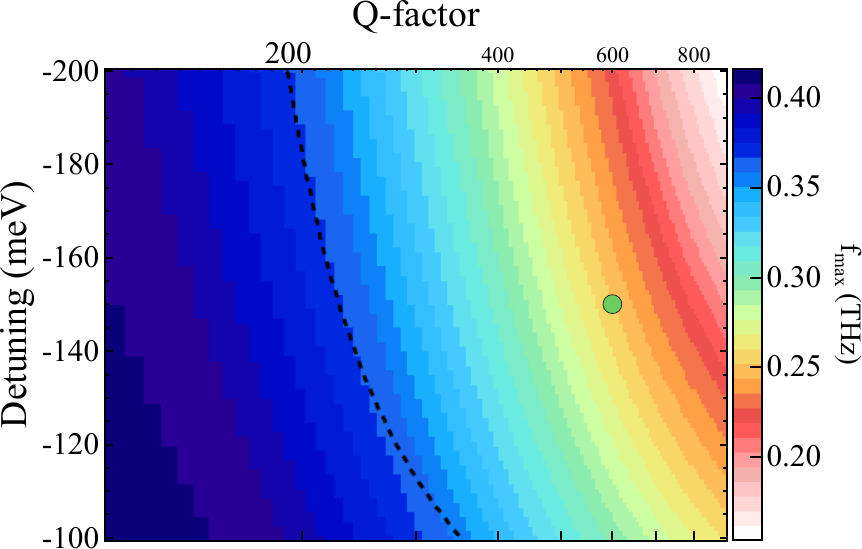}
   % \vspace*{-0.25cm}
    \caption{\label{SI:5} Optimization of the platform. Maximum clock frequency as a function of detuning and quality factor is plotted with color. The parameter space below the black dashed line is inaccessible due to optical saturation of the sample. The sample, used in this study, is marked with a green dot. } 
    %\vspace*{-0.5cm}
\end{figure}

In this Section, we provide an extended analysis of the system design optimization to maximize the maximum clock frequency. As deduced in Section I, the minimal temporal separation between the two consequent pulses that allows for correct operation of the device is of the order of 4.2 ps and is limited mostly by the ground state lifetime, $\tau_{\rm g.s.}$. We notice that both numerical and experimental study were conducted with pulses, shorter than any characteristic time of the system, therefore such limitation is intrinsic to the sample. The two parameters that influence $\tau_{\rm g.s.}$ are dependant on cavity design and not on the properties of the polymer (optimization of the molecule synthesis goes beyond the scope of this study) are the cavity quality factor, $Q$, and the exciton-photon detuning, $\delta$. We repeat the numerical time-delay scan for each point in the ($Q$, $\delta$) parameter space and plot the corresponding maximum allowed repetition rate with colour on Figure \ref{SI:5}. The set of parameters for the microcavity studied in this Letter is marked with a green dot: $Q=600$, $\delta=-0.15$. 

However, we need to estimate the applicability boundaries of our approach. The first limitation on the quality factor decrease is the loss of strong coupling regime. For the parameters, presented in Table \ref{tab:1} and the strong coupling condition, $\Omega_{R}\gg\gamma_{\rm polariton}$, we estimate that $Q$ should be above $Q_{\rm weak}\approx28$. The second limitation stems from the fact that we do not consider the total amount of unexcited molecules in the system, and practically ``inject" hot excitons in the reservoir. Such approach is applicable when we work far below optical saturation, but as we start to decrease the $Q$-factor, the pumping power, required for condensation to occur, $P_{\rm th}$, rises. At some point the required amount of hot excitons in the simulation becomes larger than half of the optically active molecules, deeming the corresponding numerical simulations unphysical. For the molecular concentration of MeLPPP and geometric parameters of the cavity, we calculate the optical saturation value of pump in the model. The line on the colourmap Fig.\ref{SI:5} where the $P_{\rm th}$ equals the optical saturation are marked with a dashed line. The region to the left of this line is, therefore, unattainable for the current pumping scheme and polymer parameters.

\end{document}